\documentclass[9pt,twoside]{article}
\usepackage{graphicx}
\usepackage{bm}
\usepackage{authblk}
\usepackage{amsmath}
\newcommand{\reffig}[1]{Fig.~\ref{#1}}

\newcommand{\refeq}[1]{Eq.~(\ref{#1})}

\title{An intuitive approach to structuring the three polarization components of light}

\author[1,2,3,*]{F. Maucher}
\author[4]{S. Skupin}
\author[1]{S. A. Gardiner}
\author[1]{I. G. Hughes}

\affil[1]{Joint Quantum Centre (JQC) Durham-Newcastle, Department of Physics, Durham University, Durham DH1 3LE, United Kingdom.}
\affil[2]{Department of Mathematical Sciences, Durham University, Durham DH1 3LE, United Kingdom.}
\affil[3]{Department of Physics and Astronomy, Aarhus University, Ny Munkegade 120, 8000 Aarhus C, Denmark}
\affil[4]{Univ.~Lyon, Universit\'e Claude Bernard Lyon 1, CNRS, Institut Lumi\`ere Mati\`ere, F-69622, Villeurbanne, France}

\affil[*]{Corresponding author: maucher@phys.au.dk}

\begin{document}

\maketitle

\begin{abstract}
This paper presents intuitive interpretations of tightly focused beams of light by drawing analogies to two-dimensional electrostatics, magnetostatics and fluid dynamics. We use a Helmholtz decomposition of the transverse polarization components in the transverse plane to introduce generalized radial and azimuthal polarization states. This reveals the interplay between transverse and longitudinal polarization components in a transparent fashion. Our approach yields a comprehensive understanding of tightly focused laser beams, which we illustrate through several insightful examples.
\end{abstract}

\section{Introduction}

Unless the beam's transverse polarization components are divergence-free in the two-dimensional transverse plane~\cite{Maucher:PRL:2018}, tightly focused light typically leads to a non-negligible longitudinal polarization component~\cite{Richards:PRS:1959,Youngworth:OE:00}, where the terms longitudinal and transverse polarization components refer to the components of the electric field that are parallel or perpendicular, respectively, to the direction of the mean Poynting flux. Having a longitudinal polarization component does not add a new degree of freedom, in the sense that all components of the electric and magnetic fields are still fixed by prescribing two polarizations in a plane. However, it is the electric field component \emph{parallel\/} to the direction of the Poynting flux that makes it somewhat special. Taking the longitudinal polarization component properly into account leads to a range of novel physical phenomena, such as a significant decrease of the focal spot size~\cite{Leuchs:OC:2000,Leuchs:PRL:2003}, the realization of so-called ``needle beams''~\cite{Chong:NatPhot:2008} and M\"{o}bius strips in the polarization of light~\cite{Bauer:Science:2015}. In the context of light-matter interaction, taking into account the effect of the longitudinal polarization component can be crucial~\cite{Quinteiro:PRL:2017}, and it may even dominate over the transverse components~\cite{Hnatovsky2011}. 

While the longitudinal polarization component and its interplay with the transverse components has attracted significant interest over the last two decades \cite{Nye:1999,Lekner:JOSA:2003}, the discussion has usually been limited to special cases assuming certain spatial symmetries, and a simple and general intuitive picture would be highly desirable. This paper aims to provide such a picture, and presents a novel approach towards an intuitive understanding of tightly focused beams by making analogies to fluid dynamics and to two-dimensional magneto- and electrostatics. For this, a two-dimensional Helmholtz decomposition of the transverse polarization components in the transverse plane~\cite{Maucher:PRL:2018} is key. 
The Helmholtz decomposition allows the generalization of the notion of ``radial'' and ``azimuthal'' polarization in the 
following way: \emph{radial polarization\/} corresponds to an electric field that is ``curl-free'' in the transverse plane 
--- which as we shall see can be interpreted as a flow solely due to sinks and sources without vorticity --- and is the 
part of the field that gives rise to the longitudinal polarization. An \emph{azimuthally polarized\/} electric field is ``divergence-free'' in the transverse plane --- analogously to a flow solely due to vorticity without any sinks or sources  --- and does not give rise to any longitudinal polarization component. This conceptually new basis turns out to be very useful for discussing Maxwell-consistent fields, and facilitates an intuitive understanding of tightly focused beams.

The paper is organized as follows. In section \ref{sec:Model} we introduce the nomenclature, present the equations of motion, and detail the connection between radially and azimuthally polarized fields, and curl- and divergence-free solutions in the transverse plane. In section \ref{sec:interpretation} we draw analogies to electrostatics, magnetostatics and fluid dynamics, hence presenting intuitive interpretations of fields that are ``divergence-free'' and ``curl-free'' in the transverse plane. Finally, in section \ref{sec:examples} we present several intuitive examples to illustrate the analogies and convey an appreciation of the concept of generalized radial and azimuthal polarization states. Starting with basic Laguerre--Gaussian (LG) modes, we show constructions of complex random and topological beams.  

\section{Model}
\label{sec:Model}
The three polarization components of a tightly focused monochromatic beam (frequency $\omega$, wavelength $\lambda$) in free space are described by 
\begin{align}
 \nabla^2{\bf E}({\bf r}_{\perp},z) + {k}^2_0{\bf E}({\bf r}_{\perp},z) &=0, \label{eq:nonparaxial}\\
 \nabla\cdot{\bf E}({\bf r}_{\perp},z) =\nabla_{\perp}\cdot{\bf E}_{\perp}+\partial_z E_z&= 0.\label{eq:divE}
\end{align}
We have introduced $k_0^2=\omega^2/c^2=(2\pi/\lambda)^2$, the transverse coordinates ${\bf r}_{\perp}=(x,y)$, and the transverse electric field components ${\bf E}_{\perp}=(E_x,E_y)$. Note that ${\bf E}$ represents the complex amplitude vector of the beam; the full electric field has an additional trivial time-dependence $\exp(-i\omega_0 t)$ that is omitted here. 

For a given amplitude vector in the focal plane\footnote{Any other $z=$ constant plane could also be used.} ${\bf E}^{\rm f}({\bf r}_{\perp})={\bf E}({\bf r}_{\perp},z=0)$, the propagation in the positive $z$ direction can be easily computed in the transverse spatial Fourier domain as
\begin{equation}\label{eq:prop}
\hat{{\bf E}}({\bf k}_{\perp},z)=\hat{\bf {E}}^{\rm f}({\bf k}_{\perp}){\rm e}^{i k_z({\bf k}_{\perp}) z},
\end{equation}
where 
$k_z({\bf k}_{\perp})=\sqrt{k_0^2-{\bf k}_{\perp}^2}$ and ${\bf k}_{\perp} = (k_x,k_y)$; the symbol~$\hat{~}$ denotes the Fourier transform with respect to ${\bf r}_{\perp}$. It is important to notice that ${\bf E}^{\rm f}$ cannot be arbitrarily prescribed in all three components. We use a two-dimensional Helmholtz-decomposition to depict the most general expression for the two transverse components, as~\cite{Maucher:PRL:2018}
\begin{equation}
{\bf E}^{\rm f}_{\perp}({\bf r}_{\perp}) =   -\begin{pmatrix}
  \partial_x V({\bf r}_{\perp})\\
  \partial_y V({\bf r}_{\perp})\\
  \end{pmatrix}+
  \begin{pmatrix} 0 & 1 \\ -1 & 0 \end{pmatrix} 
  \begin{pmatrix}   
  \partial_x W({\bf r}_{\perp})\\
  \partial_y W({\bf r}_{\perp})\\ 
  \end{pmatrix},
\label{eq:helmholtz}
\end{equation}
where $V(\bm r_\perp)$ and $W(\bm r_\perp)$ denote arbitrary (sufficiently well behaved) scalar potentials. The first term, which we denote ${\bf E}_{\perp}^{{\rm f},V}=-\nabla_{\perp} V$, corresponds to a field that is curl-free in the transverse two-dimensional plane, i.e., $\nabla_{\perp} \times {\bf E}^{{\rm f},V}_{\perp}=0$. The second term we denote ${\bf E}_{\perp}^{{\rm f},W}=(\partial_yW,-\partial_xW)$, which gives rise to a divergence-free field in the transverse two-dimensional plane,\footnote{The electric field in vacuum always fulfills $\nabla \cdot{\bf E}=0$. Hence, the restriction ``in the transverse two-dimensional plane'' is crucial in the present context.} that is, $\nabla_{\perp} \cdot {\bf E}^{{\rm f},W}_{\perp}=0$. The longitudinal component $E_z$ is therefore coupled solely to ${\bf E}_{\perp}^{{\rm f},V}$, and together with the potential $V$ it obeys a Poisson equation, 
\begin{equation}
 \Delta_{\perp} V({\bf r}_{\perp}) = \partial_z E_z|_{z=0}({\bf r}_{\perp})\label{eq:poisson}. 
\end{equation}
Because the $z$ dependence of $E_z$ is known [see Eq.~(\ref{eq:prop})], $E_z^{\rm f}$ can be readily obtained in Fourier space from
\begin{equation}
 -{\bf k}_{\perp}^2 \hat V({\bf k}_{\perp}) = i {k_z({\bf k}_{\perp})\hat{E}_z^{\rm f}({\bf k}_{\perp})} \label{eq:phi}.
\end{equation}
For given potentials $V$ and $W$ the full solution can therefore be  written in transverse Fourier space as 
\begin{equation}\label{eq:full_solution}
\hat{\bf E}({\bf k}_{\perp},z) = i\left[
  \begin{pmatrix}
  -k_x \\
  -k_y \\
  \frac{{\bf k}_{\perp}^2}{k_z({\bf k}_{\perp})}
  \end{pmatrix}\hat V({\bf k}_{\perp})+
  \begin{pmatrix}   
  k_y \\
  -k_x \\ 
  0
  \end{pmatrix}\hat W({\bf k}_{\perp})
  \right]
   {\rm e}^{i k_z({\bf k}_{\perp}) z}.
\end{equation}

Hence, the potential $V$ generates the electric field ${\bf E}^{V}$, the three vector components of which are in general nonzero. By way of contrast, the potential $W$ generates the electric field ${\bf E}^{W}$, the longitudinal component of which vanishes (${E}^{W}_z=0$). For reasons which become clear in section~\ref{sec:examples}\ref{subsec:radaz}, we will call ${\bf E}^{V}$ \emph{radially polarized},  and ${\bf E}^{W}$ \emph{azimuthally polarized}. Other polarization states imply certain conditions on the potentials  $V$ and $W$: \emph{Linear polarization}, with $\alpha E_x=(1-\alpha) E_y$ with $0\le\alpha\le1$, requires
\begin{equation} \label{eq:lin_pol}
\left[\alpha k_x-(1-\alpha) k_y \right]\hat V({\bf k}_{\perp}) = \left[\alpha k_y+(1-\alpha) k_x \right]\hat W({\bf k}_{\perp});
\end{equation}
and \emph{circular polarization}, with $E_x= \pm i E_y$, requires
\begin{equation} \label{eq:circ_pol}
\hat V({\bf k}_{\perp}) = \pm i \hat W({\bf k}_{\perp}).
\end{equation}
In this picture, the well-known fact that any nonzero linearly or circularly polarized field necessarily gives rise to a longitudinal polarization component becomes immediately clear; in these cases $V$ is nonzero and has a nontrivial spatial dependence, and hence \refeq{eq:poisson} gives rise to a nonzero $E_{z}$.

\section{Interpretations and Analogies}\label{sec:interpretation}
Let us now interpret the equations from the last section and lay out analogies to two-dimensional electro-  and magnetostatics and fluid dynamics.

\subsection{Curl-free fields produced by potential \emph{V}}
We first consider fields ${\bf E}^{V}$ that are curl-free in the transverse plane, i.e., fields obtained when $W=0$. The aim is to understand how the transverse polarization components ${\bf E}_{\perp}^{{\rm f}, V}$ relate to the longitudinal component $E_z^{\rm f}$. Equation~(\ref{eq:poisson}) represents a Poisson equation in two dimensions for the potential $V(x,y)$, the solution of which in transverse Fourier space is given by~\refeq{eq:phi}. The source term for this Poisson equation is given by $\partial_zE_z|_{z=0}$, which depends on the transverse coordinates ${\bf r}_{\perp}$ only.

The source term $\partial_zE_z|_{z=0}$ involves the longitudinal derivative, and not the component $E_z^{\rm f}$ directly. However, making use of the solution \refeq{eq:prop} in transverse Fourier space we can approximate the source term to lowest order as
\begin{equation}
 \partial_z \hat E_z|_{z=0}({\bf k}_{\perp}) = ik_z({\bf k}_{\perp}) \hat E_z^{\rm f}({\bf k}_{\perp}) \approx ik_0 \hat E_z^{\rm f} ({\bf k}_{\perp}) + \mathcal{O}\left(\frac{ {\bf k}_{\perp}^2 }{k_0^2}\right)\label{eq:lowest_order},
\end{equation}
thereby disposing of this derivative term. Equation~(\ref{eq:lowest_order}) is of course only a rough estimate, with an error that grows with the degree of the light's non-paraxiality. It nevertheless gives an idea of the qualitative behaviour of $E_z^{\rm f}$, as will be illustrated in section \ref{sec:examples}.

To see the analogy with two-dimensional electrostatics, we must simply identify $-\partial_zE_z|_{z=0}$ with a charge density $\rho(x,y)$. 
The transverse polarization components ${\bf E}_{\perp}^{{\rm f}, V}$ then follow from the same equations as the two-dimensional electrostatic field ${\bf E}_{\perp}^{\rm s}(x,y)$, as summarized in table~\ref{table:analogy_electro}. 

\begin{table}[ht]
\centering
\caption{\bf Analogy with electrostatics in two dimensions.}
\begin{tabular}{||c c c||}
\hline
\multicolumn{2}{||c}{Electrostatics (2D)} & $W=0$ solutions\\
\hline\hline
 Charge: & $\rho(x,y)$ & $-\partial_zE_z|_{z=0}$  \\ 
 Curl-free: & $\nabla_{\perp} \times {\bf E}_{\perp}^{\rm s}=0$ & $\nabla_{\perp}\times {\bf E}_{\perp}^{{\rm f}, V}=0$  \\ 
 Field: & ${\bf E}_{\perp}^{\rm s}=-\nabla_{\perp} \phi$ & ${\bf E}_{\perp}^{{\rm f}, V}=-\nabla_{\perp} V$  \\ 
 Poisson-eq.: & $ \Delta_{\perp}\phi=-\rho(x,y)$ & $\Delta_{\perp} V=\partial_zE_z|_{z=0}$ \\
 \hline
\end{tabular}
  \label{table:analogy_electro}
\end{table}

There is, however, a subtlety related to this analogy requiring of comment. The optical field amplitude ${\bf E}$ is complex-valued, and the full electric field has a temporal dependence that is hidden in this representation, unlike the two-dimensional electrostatic field which is real-valued and time-independent. Thus, in general, we must separate the real and imaginary parts of ${\bf E}$ and solve two ``electrostatic'' problems. 

\subsection{Divergence-free fields produced by potential \emph{W}}
Here we are interested in solutions ${\bf E}^W$ that are divergence-free in the transverse plane ($V=0$, and consequently $E_z^W=0$). 
The potential $W$ can be interpreted as the longitudinal component of a vector potential, i.e., ${\bf E}^{{\rm f}, W}=\nabla\times W {\bm e}_z$. Then, by taking the curl of this equation, we get
\begin{equation}\label{eq:poisson_W}
 \Delta_{\perp} W({\bf r}_{\perp}) = \partial_yE_x^{\rm f}({\bf r}_{\perp}) - \partial_xE_y^{\rm f}({\bf r}_{\perp}),
\end{equation}
a Poisson equation for the potential $W$.

This leads to a straightforward analogy in \emph{magnetostatics}\footnote{We follow the common terminology which labels magnetic fields of stationary currents as \emph{magnetostatic}, even though time inversion symmetry is violated.}. If we identify the potential $W$ with the only nonzero component $A_z^{\rm s}$ of a magnetic vector potential, 
\begin{equation}
 {\bm A}^{\rm s}({\bf r}_{\perp}) = A_z^{\rm s}({\bf r}_{\perp}) {\bm e}_z,
\end{equation}
the induced static magnetic field ${\bm B}^{\rm s}=\nabla \times {\bm A}^{\rm s}$
can be associated with ${\bm E}^{{\rm f}, W}$. The static current density follows from $\nabla \times {\bm B}^{\rm s} = \mu_0 {\bm J}^{\rm s}$, and flows perpendicular to the $(x,y)$ plane under consideration. Hence, $J^{\rm s}_z$ is the only nonzero component, and corresponds to the negative source term in \refeq{eq:poisson_W}. The complete magnetostatic analogy is summarized in table~\ref{table:analogy_magnetostatics}.

\begin{table}[ht]
\centering
\caption{\bf Analogy with magnetostatics in two dimensions.}
\begin{tabular}{||c c c||}
\hline 
\multicolumn{2}{||c}{Magnetostatics with ${\bm J}^{\rm s}=J^{\rm s}_z({\bf r}_{\perp}) {\bm e}_z$} &$ V=0$ solutions\\
\hline\hline
 Current: & $\mu_0J^{\rm s}_z=\partial_yB_x^{\rm s}-\partial_xB_y^{\rm s}$ &  $\partial_xE_y^{\rm f}-\partial_yE_x^{\rm f}$ \\ 
 Div.-free: & $\nabla_{\perp} \cdot {\bf B}_{\perp}^{\rm s}=0$ & $\nabla_{\perp} \cdot {\bf E}_{\perp} ^{{\rm f}, W}=0$  \\ 
 Field: & ${\bf B}_{\perp}^{\rm s}=(\partial_y A^{\rm s}_z,-\partial_x A^{\rm s}_z)$ & ${\bf E}_{\perp}^{{\rm f}, W}=(\partial_y W,-\partial_x W)$  \\
 Poisson-eq.: & $\Delta_{\perp} A^{\rm s}_z = -\mu_0J^{\rm s}_z$ & $\Delta W_{\perp} = \partial_yE_x^{\rm f}-\partial_xE_y^{\rm f}$  \\
 \hline
\end{tabular}
  \label{table:analogy_magnetostatics}
\end{table}

A similar remark as given above concerning the complex-valued optical field amplitude applies, in that two real-valued magnetostatic problems may need to be solved to describe one curl-free optical field.

\subsection{Fields produced by both potentials \emph{V} and \emph{W}}
In the most general situation, both the $V$ and $W$ potentials are nonzero. In this case the optical field ${\bm E}^{\rm f}= {\bm E}^{{\rm f},V}+{\bm E}^{{\rm f},W}$ has a perfect analogy in fluid dynamics~\cite{fluid_dynamics}. For a fluid with density $\varrho$ and flow velocity field ${\bf u}$, conservation of mass dictates the continuity equation
\begin{equation}
\partial_t  \varrho({\bf r},t) + \nabla \cdot \left[ \varrho({\bf r},t) {\bf u}({\bf r},t) \right] =0.
\end{equation}
We consider a layer of this fluid in a $z=$constant plane, e.g. $z=0$, with in- and outflow $\varrho u_z$ into the layer. We denote the flow velocity field in this layer ${\bf u}^{\rm l}({\bf r}_{\perp}) = {\bf u}({\bf r}_{\perp},z=0)$. In the stationary situation ($\partial_t\varrho=0$), and furthermore assuming homogenity and incompressibility ($\nabla\varrho=0$), we find that 
\begin{equation}\label{eq:nalbau}
 \nabla_{\perp} \cdot {\bf u}^{\rm l}_{\perp}({\bf r}_{\perp}) =  - \partial_z u_{z}|_{z=0}({\bf r}_{\perp}).
\end{equation}
Equation~(\ref{eq:nalbau}) is identical to \refeq{eq:divE} if we identify ${\bf u}^{\rm l}_{\perp}$ with ${\bm E}^{\rm f}_{\perp}$,
and we write ${\bf u}_{\perp}^{\rm l}={\bf u}_{\perp}^{\rm l, cf}+{\bf u}_{\perp}^{\rm l, df}$, where 
${\bf u}_{\perp}^{\rm l,cf}$ is curl-free in the $z=0$ plane ($\nabla_{\perp} \times {\bf u}_{\perp}^{\rm l, cf}=0$),
and ${\bf u}_{\perp}^{\rm l,df}$ divergence-free ($\nabla_{\perp} \cdot {\bf u}_{\perp}^{\rm l, df}=0$). 

The analogy between ${\bf u}_{\perp}^{\rm l,cf}$ and ${\bm E}^{{\rm f},V}$ follows immediately: the curl-free velocity field ${\bf u}^{\rm l,cf}_\perp$ can be written as the negative gradient of the so-called velocity potential $\phi$, and the spatial in- and outflow rate  $\partial_z u_{z}|_{z=0}$ acts as a source term in the Poisson equation for the velocity potential, as shown in table~\ref{table:analogy_fluid_cf}.

\begin{table}[ht]
\centering
\caption{\bf Analogy with curl-free velocity field flow.}
\begin{tabular}{||c c c||}
\hline
\multicolumn{2}{||c}{Fluid dynamics (2D)} & $W=0$ solutions\\
\hline\hline
 Sources \& sinks: & $\partial_zu_z|_{z=0}$ & $\partial_zE_z|_{z=0}$  \\ 
 Curl-free: & $\nabla_{\perp} \times {\bf u}_{\perp}^{\rm l,cf}=0$ & $\nabla_{\perp}\times {\bf E}_{\perp}^{\rm f, V}=0$  \\ 
 Field: & ${\bf u}_{\perp}^{\rm l,cf}=-\nabla_{\perp} \phi$ & ${\bf E}_{\perp}^{\rm f, V}=-\nabla_{\perp} V$  \\ 
 Poisson-eq.: & $ \Delta_{\perp}\phi=\partial_zu_z|_{z=0}$ & $\Delta_{\perp} V=\partial_zE_z|_{z=0}$ \\
 \hline
\end{tabular}
  \label{table:analogy_fluid_cf}
\end{table}

Following~\cite{fluid_dynamics}, let us now turn towards the divergence-free (two-dimensional) velocity field ${\bf u}_{\perp}^{\rm l, df}$, which obeys the continuity equation
\begin{equation}
 \partial_x u_x^{\rm l, df}({\bf r}_{\perp}) + \partial_y u_y^{\rm l, df}({\bf r}_{\perp}) = 0.
 \label{eq:divu2d}
\end{equation}
This implies that the differential $d\psi=u_x^{\rm l, df}dy-u_y^{\rm l, df}dx$ is exact, and the scalar \emph{stream function\/} $\psi$  can be found (up to a constant) as line integral from some reference point $\mathcal{O}$ to a given point $\mathcal{P}$, 
\begin{equation}\label{eq:lineintfluid}
 \psi(\mathcal{P})-\psi(\mathcal{O})=\int_{\mathcal{O}}^{\mathcal{P}} d\psi=\int_{\mathcal{O}}^{\mathcal{P}} u^{\rm l, df}_xdy-u^{\rm l, df}_ydx.
\end{equation}
In particular any integration curve joining the two points yields the same result.\footnote{We consider a simply connected region.} In fluid dynamics, the stream function characterizes the flow velocity quite intuitively, as already suggested by its name. The ``flux'' across a closed curve, i.e. $\mathcal{P}=\mathcal{O}$, is zero. 
Since the ``flux'' across any curve joining the two points $\mathcal{P}$ and $\mathcal{O}$ depends only on the values of $\psi$ at these two points, 
it is clear that $\psi$ is constant along a streamline. 

In fluid dynamics, one commonly defines the \emph{vorticity\/} as 
\begin{equation}
 {\bm\omega}=\nabla\times {\bf u}_{\perp}^{\rm l}=(\partial_xu_y^{\rm l}-\partial_yu_x^{\rm l}){\bm e}_z=\Omega{\bm e}_z.
\end{equation}
The vorticity vector is oriented perpendicular to our plane of interest, and its non-zero component $\Omega$, together with the stream function $\psi$,  obeys a Poisson equation:
\begin{equation}
 \Delta_{\perp} \psi({\bf r}_{\perp}) = -\Omega({\bf r}_{\perp}). 
\end{equation}
Hence, ${\bm E}_{\perp}^{{\rm f}, W}$ can be interpreted as the divergence-free velocity field ${\bf u}_{\perp}^{\rm l, df}$ of a two-dimensional incompressible fluid, which is composed of vortices only and does not contain sources or sinks. The potential $W$ must be associated with the fluid stream function $\psi$ determined by the longitudinal component $\Omega$ of the {\em vorticity} ${\bm \omega}$. The full analogy is summarized in table~\ref{table:analogy_fluid}.

\begin{table}[ht]
\centering
\caption{\bf Analogy with divergence-free velocity field flow.}
\begin{tabular}{||c c c||}
\hline 
\multicolumn{2}{||c}{Fluid dynamics (2D)} &$ V=0$ solutions\\
\hline\hline
 Vorticity: & $\Omega = \partial_xu_y^{\rm l}-\partial_yu_x^{\rm l}$ &  $\Omega = \partial_xE_y^{\rm f}-\partial_yE_x^{\rm f}$ \\ 
 Div.-free: & $\nabla_{\perp} \cdot {\bf u}_{\perp}^{\rm l, df}=0$ & $\nabla_{\perp} \cdot {\bf E}_{\perp} ^{{\rm f}, W}=0$  \\ 
 Field: & ${\bf u}_{\perp}^{\rm l, df}=(\partial_y \psi,-\partial_x \psi)$ & ${\bf E}_{\perp}^{{\rm f}, W}=(\partial_y W,-\partial_x W)$  \\
 Poisson-eq.: & $\Delta \psi = -\Omega$ & $\Delta W = \partial_yE_x^{\rm f}-\partial_xE_y^{\rm f}$  \\
 \hline
\end{tabular}
  \label{table:analogy_fluid}
\end{table}

In summary, we note that two-dimensional incompressible fluid dynamics permits coverage of the cases of both curl- and divergence-free fields ($V=0$ and $W=0$, respectively), as shown in tables~\ref{table:analogy_fluid_cf} and \ref{table:analogy_fluid},  and thus represents a unified analogy. One must bear in mind, however, that in fluid dynamics the flow velocity field ${\bf u}$ is real valued, and so in general two fluid problems are necessary to represent one complex optical field.

\section{Examples}\label{sec:examples}
We now present several intuitive examples to illustrate our findings, highlighting the concept of generalized radial and azimuthal polarization, and exploiting the aforementioned analogies to intuitively construct Maxwell-consistent vector beams. 

There are two general remarks to be made before prescribing any potential or field components. Firstly, the relevant quantities must not contain any evanescent amplitudes, that is, in the transverse Fourier domain ${\bf k}_\perp$ all fields and potentials must be zero for ${\bf k}_{\perp}^2\ge k_0^2$. Otherwise, \refeq{eq:prop} gives an exponentially growing solution for negative $z$, which renders it unphysical in the bulk. Therefore, in the examples that follow we systematically apply the filter function 
\begin{equation}\label{eq:Hk0}
H_{k_0}(\bm k_\perp)=\begin{cases}
 \exp \left[{-\frac{1}{ 2\lambda^2\left(\sqrt{\bm k_\perp^2}-k_0\right)^2}} \right] &\text{for } \bm k_\perp^2<k_0^2 \\
  0 &\text{for } \bm k_\perp^2\geq k_0^2
 \end{cases}
\end{equation}
to these quantities. Secondly, Eq.~(\ref{eq:divE}) implies that ${\hat{E}_z^{\rm f}}({\bf k}_{\perp}=0)=0$ for solutions propagating in the $z$ direction. This is automatically fulfilled when ${{E}_z^{\rm f}}$ (or $\partial_z {E}_z|_{z=0}$) is computed from a given potential $V$. However, if ${{E}_z^{\rm f}}$ (or $\partial_z {E}_z|_{z=0}$) is prescribed, special care must be taken. A suitable filter function is then 
\begin{equation}\label{eq:H0}
H_{0}(\bm k_\perp)=1-e^{-(3\lambda{\bm k_\perp})^2},
\end{equation}
which we have applied in the examples below when needed.

\subsection{Radial and azimuthal polarization}\label{subsec:radaz}
Let us first consider the simplest examples of bell-shaped $V$ and $W$ potentials, producing classic radially and azimuthally polarized vector beams, respectively. In terms of LG profiles, we therefore take ${\rm LG}^\sigma_{00}({\bf r}_{\perp})$, i.e., a simple Gaussian, fixing its width to $\sigma=\lambda/2$. As explained above, we must filter the potentials in the transverse Fourier domain by multiplying them by $H_{k_0}$ in order to remove evanescent waves. 

\subsubsection{Gaussian ``electrostatic potential'' V}

\begin{figure}[ht]
 \includegraphics[width=.9\columnwidth]{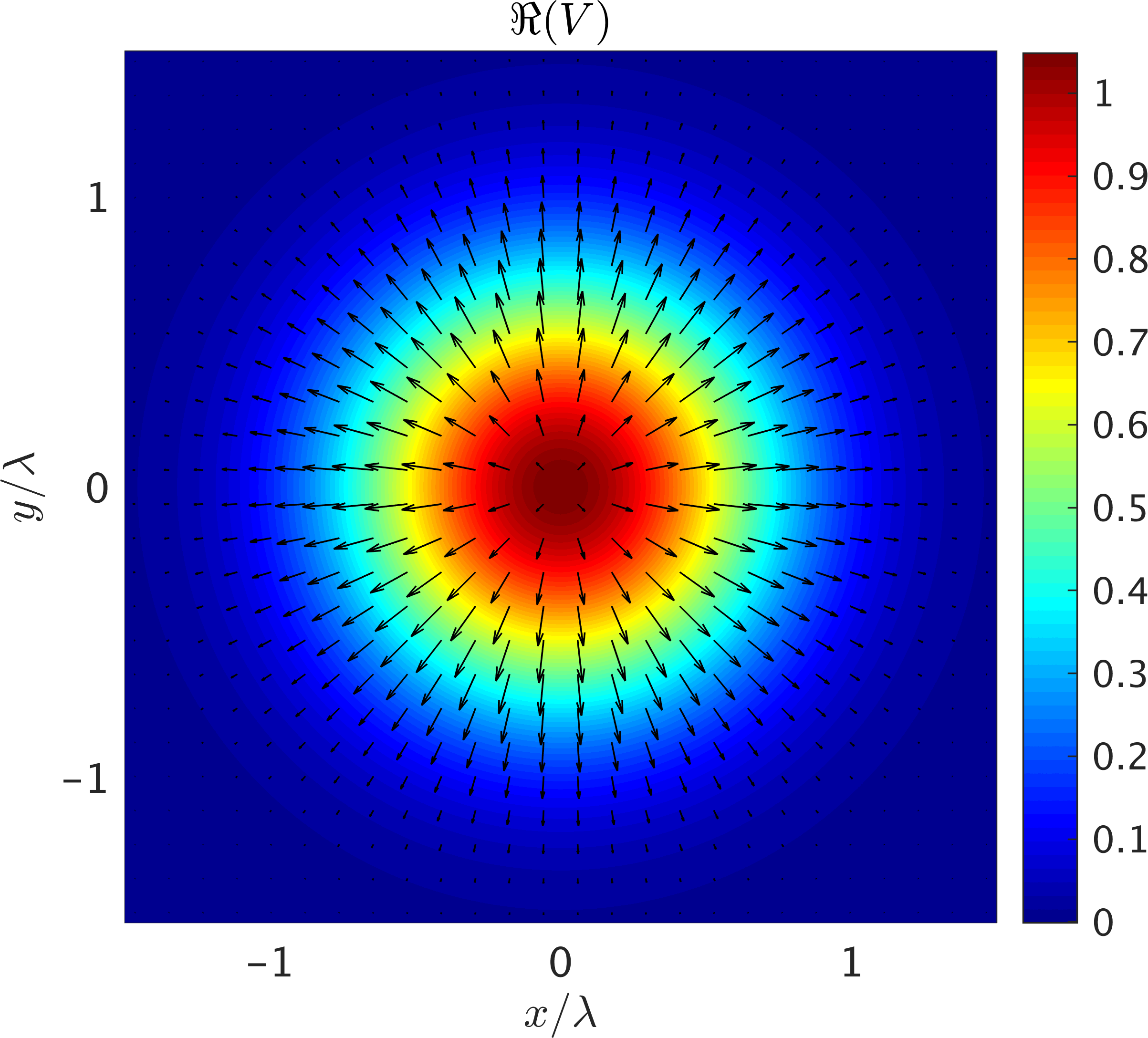}
 \includegraphics[width=.9\columnwidth]{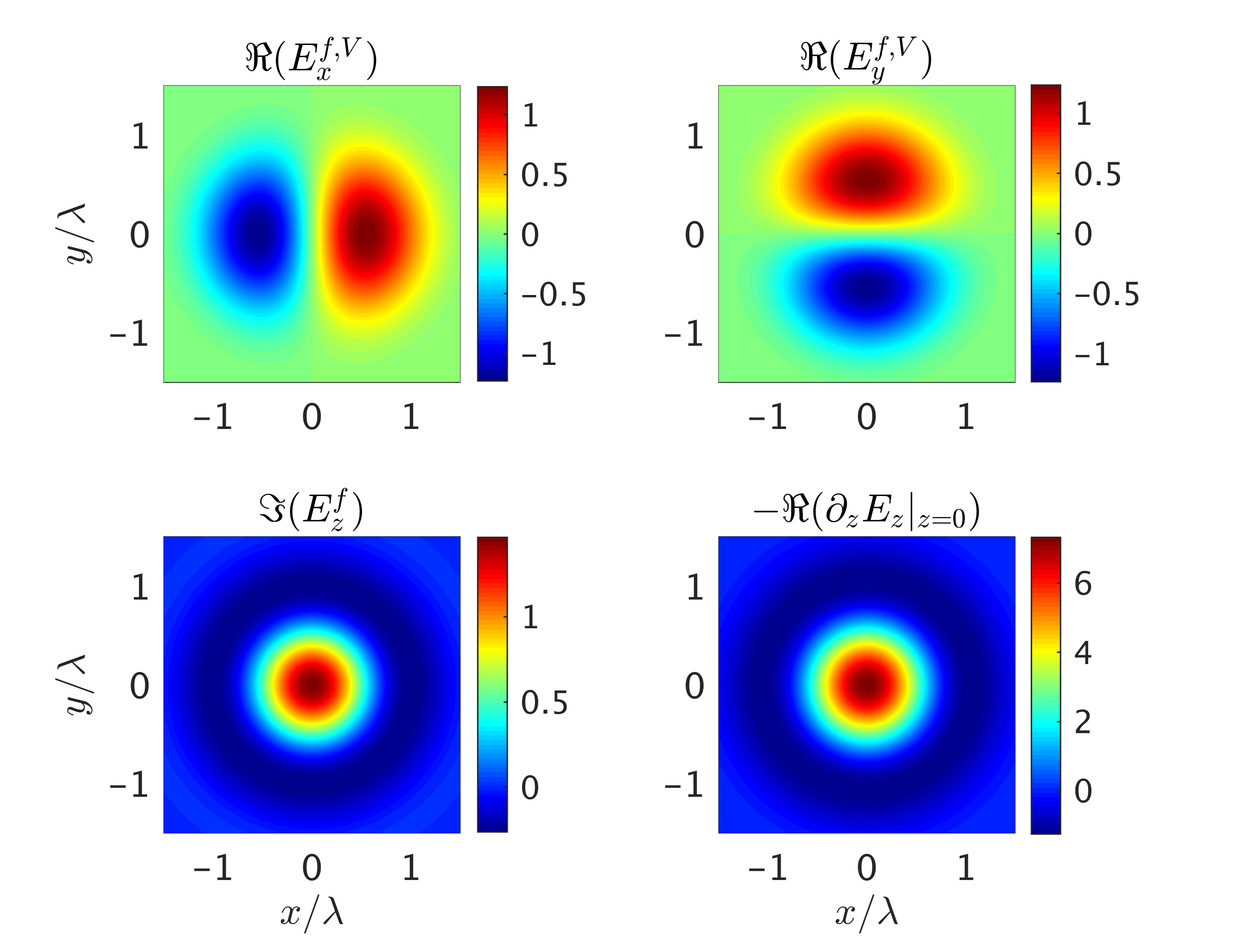} 
 \caption{A real-valued Gaussian potential $V \propto \exp(-2{\bm r}^2_{\perp}/\lambda^2)$ produces a radially polarized beam. The transverse electric field ${\bm E}^{{\rm f},V}_{\perp}$ is indicated by arrows. The induced field components and the "charge density" $-\partial_zE_z|_{z=0}$ are shown below.}
\label{fig:gaussian_V}
\end{figure}

Given that we use the constraint $W=0$, and $V$ is chosen to be a real-valued function, we require only the real parts of ${\bm E}_{\perp}^{{\rm f},V}$ and $\partial_zE_z|_{z=0}$, and the imaginary part of $E_z^{\rm f}$. We therefore depict these components only in Fig.~\ref{fig:gaussian_V}, which displays the resulting radially polarized beam. It follows straightforwardly that we get two dipole-like light distributions in the transverse polarization components from such a bell-shaped ``electrostatic potential'' $V$. Going further with this analogy, the ``charge density'' 
$-\partial_zE_z|_{z=0}\propto(1-2{\bm r}^2_{\perp}/\lambda^2)\exp(-2{\bm r}^2_{\perp}/\lambda^2)$ inducing such a potential consists of a positive hump and a negative ring. The shape of this``charge density'' is very close to the longitudinal component $E_z^{\rm f}$, which justifies the estimation given by \refeq{eq:lowest_order}.

\subsubsection{Gaussian "magnetic vector potential" W}
\begin{figure}[ht]
 \includegraphics[width=0.9\columnwidth]{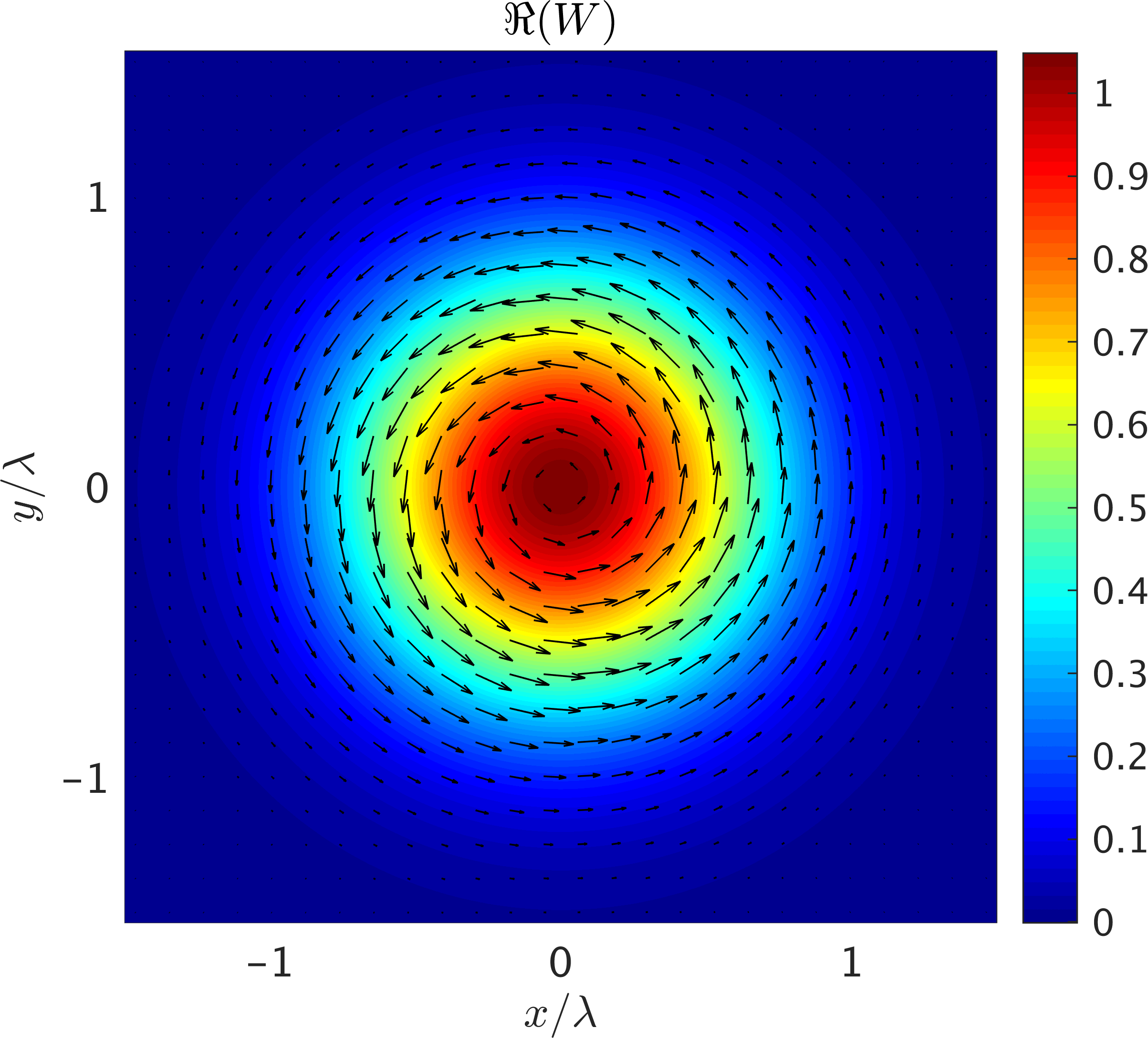} 
 \includegraphics[width=0.9\columnwidth]{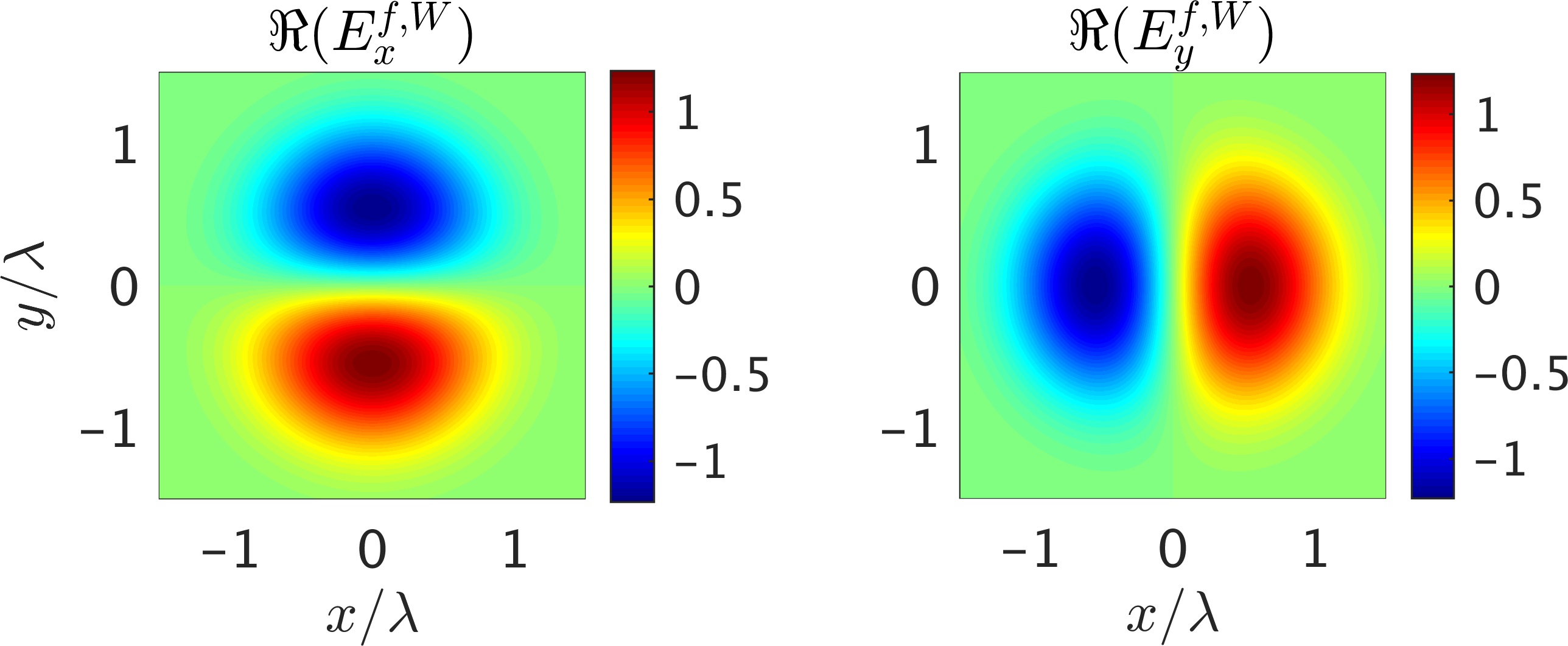} 
 \caption{A real-valued Gaussian potential $W \propto \exp(-2{\bm r}^2_{\perp}/\lambda^2)$ produces an azimuthally polarized beam. The purely transverse electric field ${\bm E}^{{\rm f},W}_{\perp}$ is indicated by arrows. The induced field components are shown below (${E}_z^{\rm f}=0$).}
\label{fig:gaussian_W}
\end{figure}

In order to produce the azimuthally polarized vector beam, we simply let $W$ be the real-valued Gaussian, and set $V=0$. Figure ~\ref{fig:gaussian_W} shows that the vector field ${\bf E}_{\perp}^{{\rm f},W}$ is tangential to the contour lines of $W$. This is exactly what one would expect from the magnetostatic analogy, where $W$ corresponds to the longitudinal component $A_z^{\rm s}$ of the magnetic vector potential. We do not depict the static current density $J_z^{\rm s}$ that would generate such a vector potential and respective magnetic field. However, its profile has exactly the same form as the ``charge density'' $-\partial_zE_z|_{z=0}$ in Fig.~\ref{fig:gaussian_V}: a positive bell-shaped current density at the center, surrounded by a negative ring-like return current.

\subsubsection{Circularly polarized vortex beam}
\begin{figure}[ht]
\includegraphics[width=0.9\columnwidth]{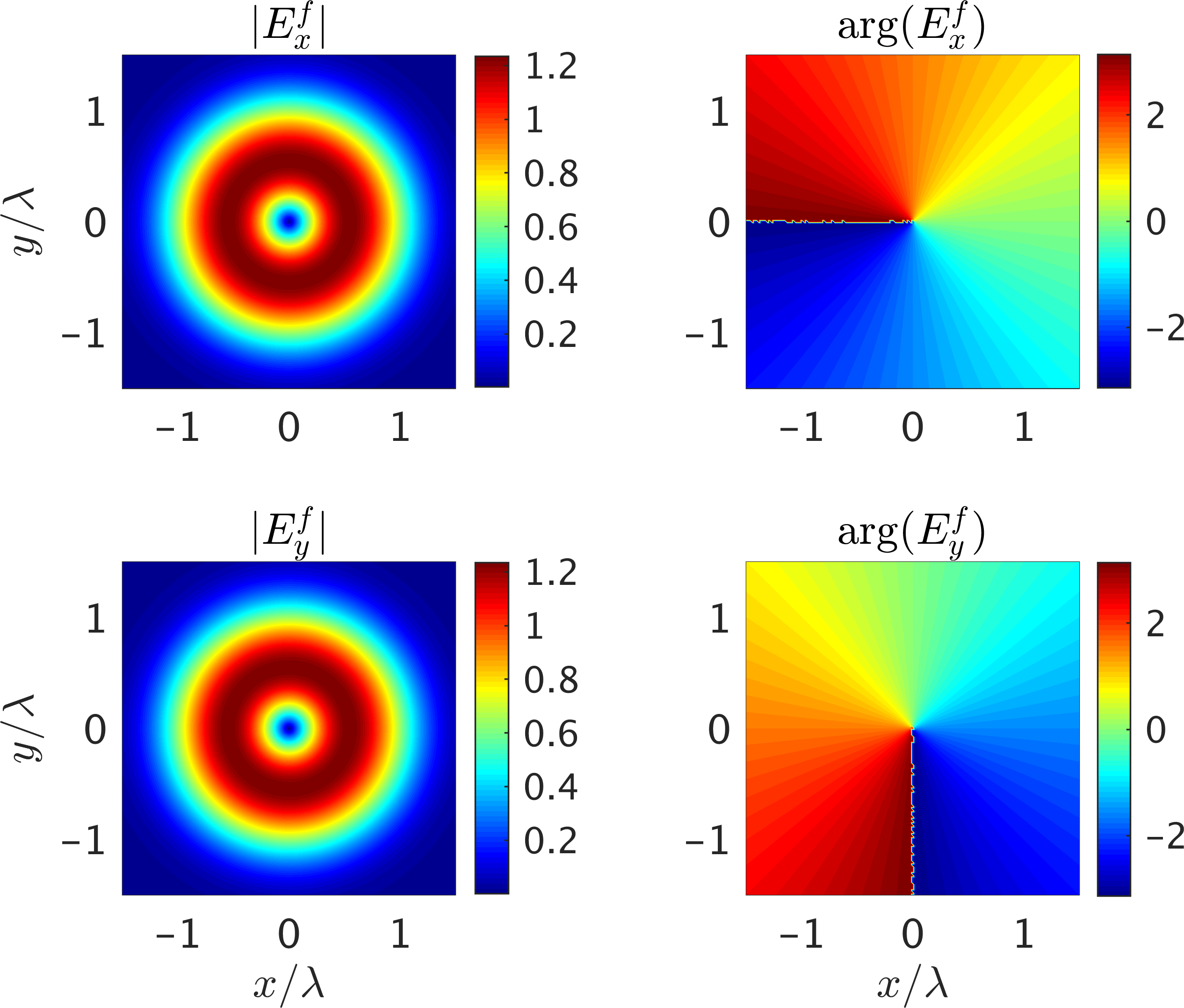} 
\caption{Transverse field components of a circularly polarized single charge vortex beam produced by complex superposition of Figs~\ref{fig:gaussian_V} and \ref{fig:gaussian_W}, i.e., $V = iW \propto \exp(-2{\bm r}^2_{\perp}/\lambda^2)$.}
\label{fig:gaussian_circ}
\end{figure}

In section~\ref{sec:Model}, we discussed how to choose the potentials $V$ and $W$ such that the transverse fields are circularly polarized, namely $\pm iW=V$. Hence, complex superposition of the fields shown in Figs.~\ref{fig:gaussian_V} and \ref{fig:gaussian_W} should give a circularly polarized beam. In Fig.~\ref{fig:gaussian_circ} we display the resulting transverse field components $E_x^{\rm f}$ and $E_y^{\rm f}$, which are indeed two singly charged vortices. The corresponding longitudinal polarization component $E_z^{\rm f}$ is, as expected, the same as that shown in Fig.~\ref{fig:gaussian_V}.

This rather straightforward construction of circularly polarized beams works for any pair of generalized radially and azimuthally polarized beams with $\pm iW=V$. The motivation for this consideration stems from the fact that in such an arrangement 
one has a beam that is both linearly (the longitudinal polarization component) as well as circularly polarized (the transverse polarization 
components). This could play an important role for imprinting structures onto matter, since the longitudinal polarization component could drive a different transition to the transverse polarization component. 

\subsection{Longitudinal vortex beam} \label{sec:vortex}
\begin{figure}[ht]
 \includegraphics[width=.9\columnwidth]{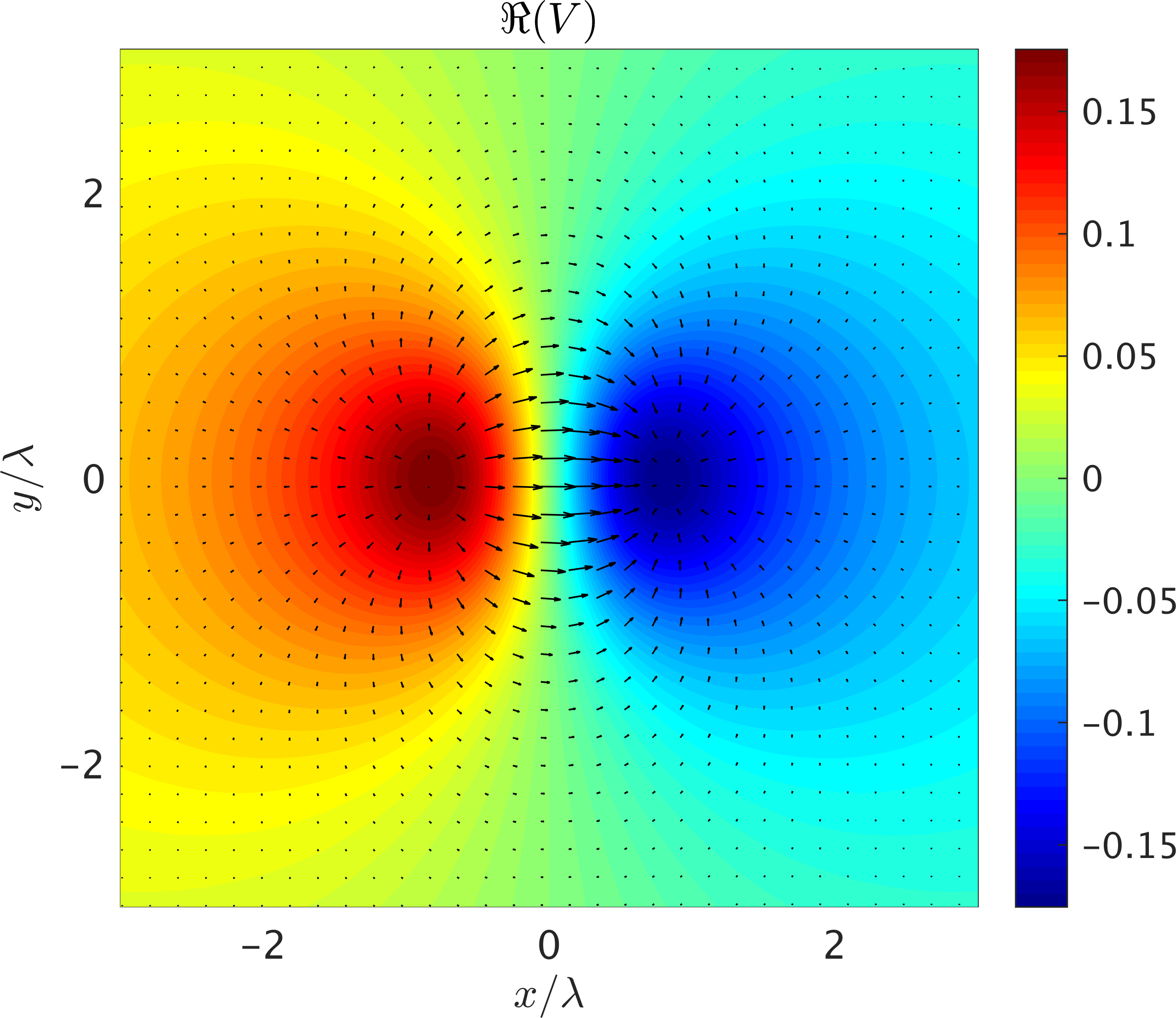}
 \includegraphics[width=.9\columnwidth]{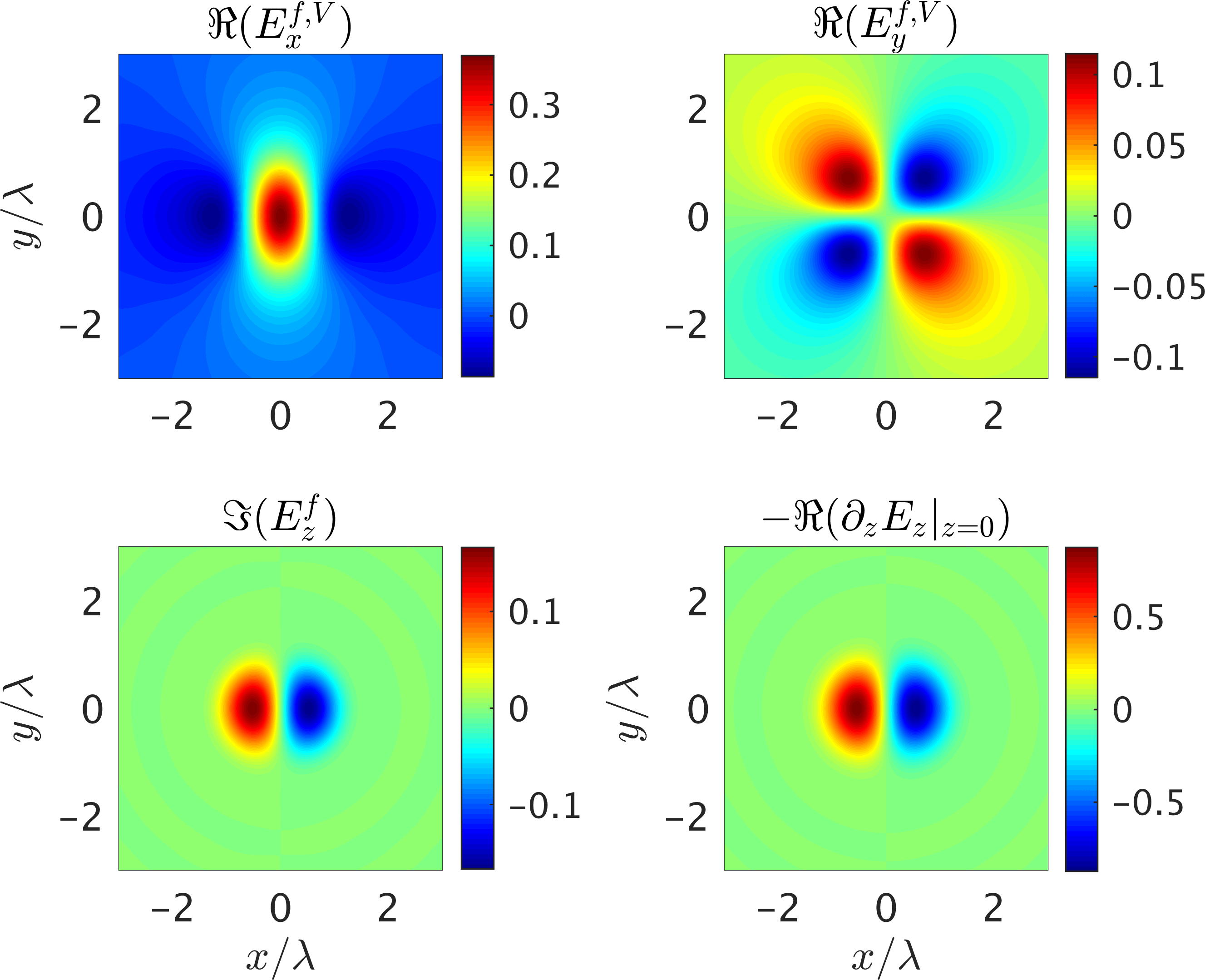}
 \caption{A real-valued dipole is prescribed in $\partial_z E_z|_{z=0}$, and the resulting potential $V=\phi^{\rm dipole}$ (see text for details) is used to compute the polarization components.}
\label{fig:interpretation_dipole}
\end{figure}

Let us now use our previous findings to construct a beam featuring a singly charged vortex in its longitudinal component $E_z^{\rm f}$. Because such a vortex is composed of two orthogonal dipoles in the real and imaginary parts, we may start with a dipole. Moreover, we want to make use of the fluid dynamics analogy, and prescribe a real-valued dipole in the ``spatial in- and outflow rate,'' $\partial_zE_z|_{z=0}\propto x\exp(-2{\bm r}^2_{\perp}/\lambda^2)$. The induced velocity potential is easily obtained in the transverse Fourier domain as 
\begin{equation}
\hat\phi^{\rm dipole} \propto H_{k_0} k_x \exp(-\lambda^2{\bm k}^2_{\perp}/8)/{\bm k}^2_{\perp},
\end{equation}
where we apply the filter function $H_{k_0}$ to remove any evanescent waves. The resulting potential in position space is shown in Fig.~\ref{fig:interpretation_dipole}, together with all polarization components. In terms of the fluid dynamics analogy, where we interpret the transverse field as the flow velocity, a very intuitive picture arises: peak and trough of the dipole in $\partial_zE_z|_{z=0}$ act like source and sink for the ``flow.'' With the approximation~\refeq{eq:lowest_order}, even the longitudinal component $E_z^{\rm f}$ by itself can be identified with the source of the ``transverse flow.''

In principle, we can now construct a beam with the desired singly charged vortex in its longitudinal component by employing the potential $V(x,y) = \phi^{\rm dipole}(x,y) \pm i \phi^{\rm dipole}(y,x)$, i.e., by adding $\phi^{\rm dipole}$ rotated by $\pm \pi/2$ as the imaginary part of $V$. The resulting beam is radially polarized in the generalized sense introduced earlier, because $W=0$. It is, however, not the easiest realization of a longitudinal vortex, as we will see below.

\begin{figure}[ht]
 \includegraphics[width=.9\columnwidth]{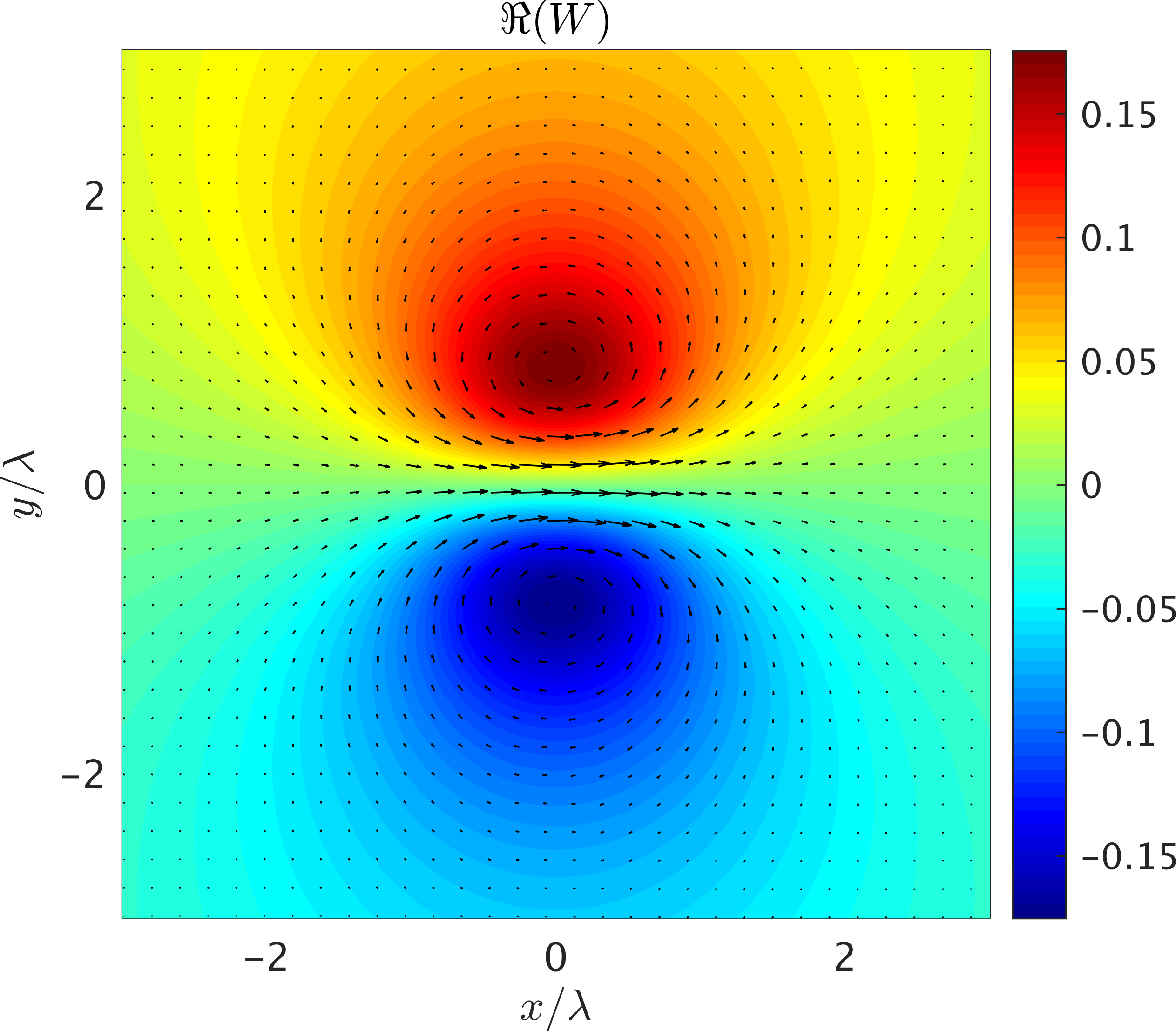}
 \includegraphics[width=.9\columnwidth]{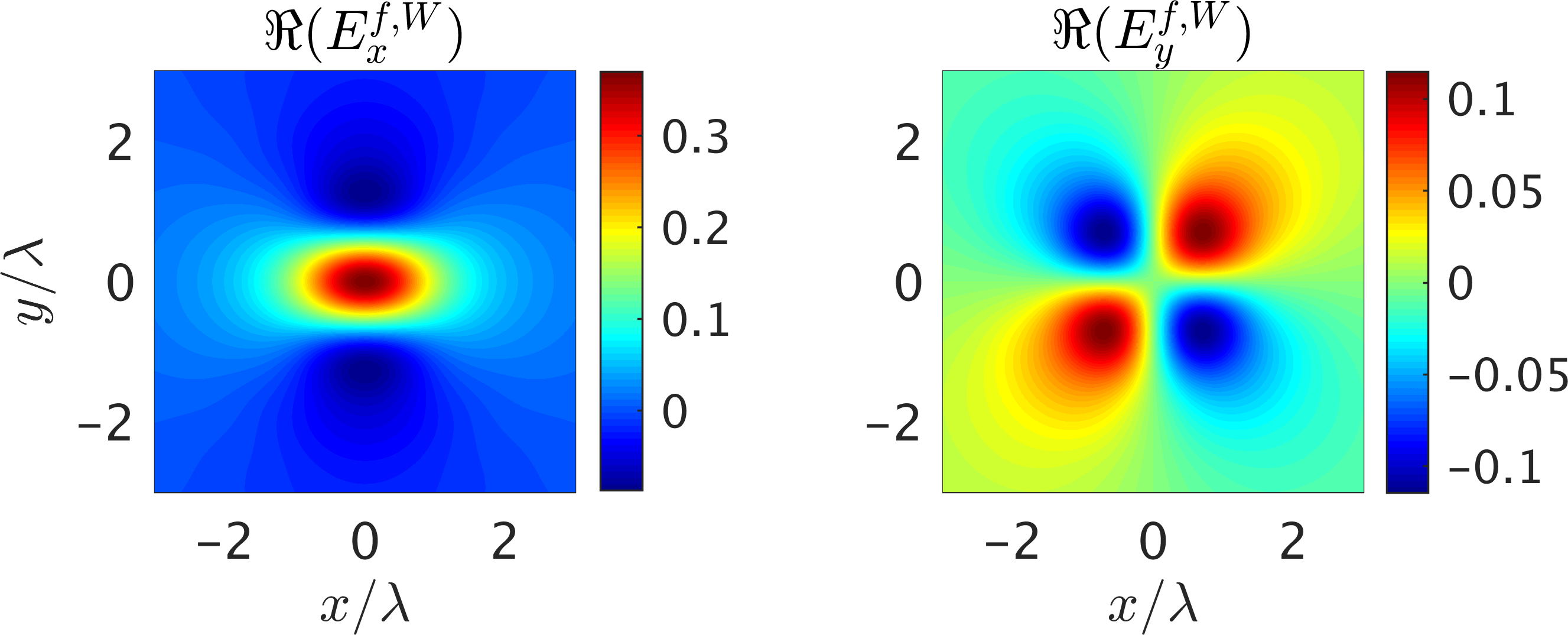}
 \caption{The real-valued potential $W(x,y) = - \phi^{\rm dipole}(y,x)$ (see text for details) produces $E_y^{{\rm f},W}=-E_y^{{\rm f},V}$ of Fig.~\ref{fig:interpretation_dipole}. }
\label{fig:dipole_W}
\end{figure}

The transverse polarization components $E_x^{{\rm f},V}$ and $E_y^{{\rm f},V}$ shown in Fig.~\ref{fig:interpretation_dipole} appear to be relatively intricate, and so the natural question arises whether it is possible to simplify the ``transverse flow'' by adding a divergence-free velocity field through an appropriate stream function $\psi$. The main flow is clearly in the positive $x$ direction, and so we may choose the stream function $\psi$ such that the $y$ component of the total flow velocity is zero. Equation~(\ref{eq:lin_pol}) tells us that $\hat \psi = - k_y \hat \phi^{\rm dipole} / k_x$ will suffice, which translates into $\psi(x,y) = - \phi^{\rm dipole}(y,x)$ in position space. And indeed, Fig.~\ref{fig:dipole_W} confirms that $E_y^{{\rm f},W}=-E_y^{{\rm f},V}$, i.e., the total ``transverse flow'' is parallel to ${\bm e}_x$.

\begin{figure}[ht]
 \includegraphics[width=.9\columnwidth]{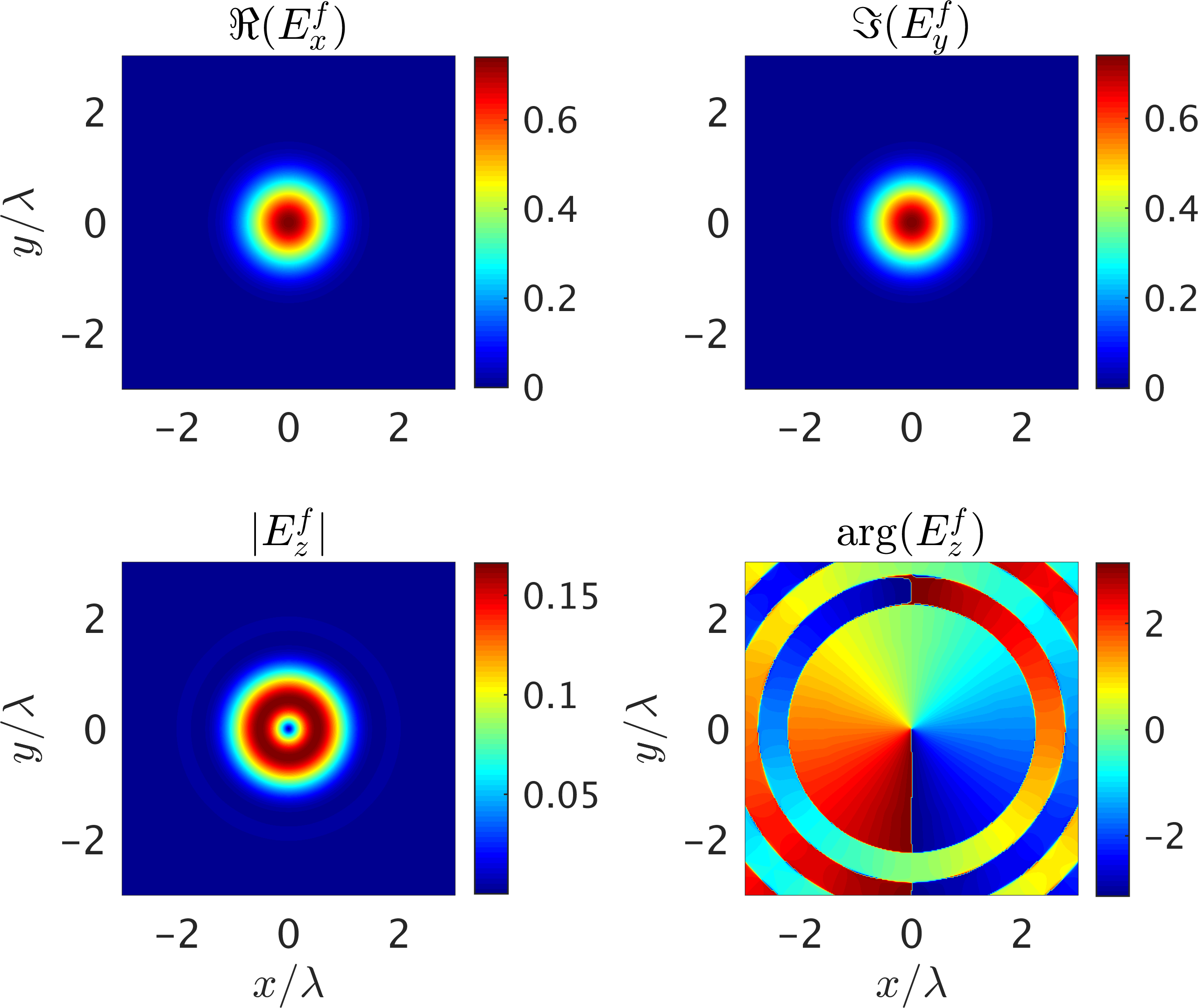}
  \caption{A bell-shaped circularly polarized transverse field produces a 
 singly charged vortex in $E_z^{\rm f}$. The off-axis sign flips in the phase plot are due to the filter function $H_{k_0}$ (see text).}
\label{fig:single_vortex}
\end{figure}

Hence, the potentials $V(x,y) = \phi^{\rm dipole}(x,y) \pm i \phi^{\rm dipole}(y,x)$ and $W(x,y) = - \phi^{\rm dipole}(y,x) \pm i \phi^{\rm dipole}(x,y)$ produce a longitudinal vortex as well, but with much simpler transverse field components. Figure~\ref{fig:single_vortex} shows the resulting optical fields (for the $+$ sign), and it even turns out that transverse components are bell-shaped. Moreover, the transverse field is circularly polarized, as $W=iV$. We have therefore constructed the solution reported in~\cite{Nieminen:JOSA:2008}, where a longitudinal vortex was achieved by tightly focusing a circularly polarized bell-shaped light distribution.

It is important to note that the choice of the ``stream function'' $W$ is the degree of freedom one has when only the longitudinal polarization component is prescribed. In the present example, we used $W$ to simplify the transverse polarization components. In section \ref{sec:examples}.\ref{sec:knots}, we will demonstrate that $W$ can be used to engineer certain topological properties of the transverse fields. 

\subsection{Random beam}
\begin{figure}[ht]
 \includegraphics[width=.9\columnwidth]{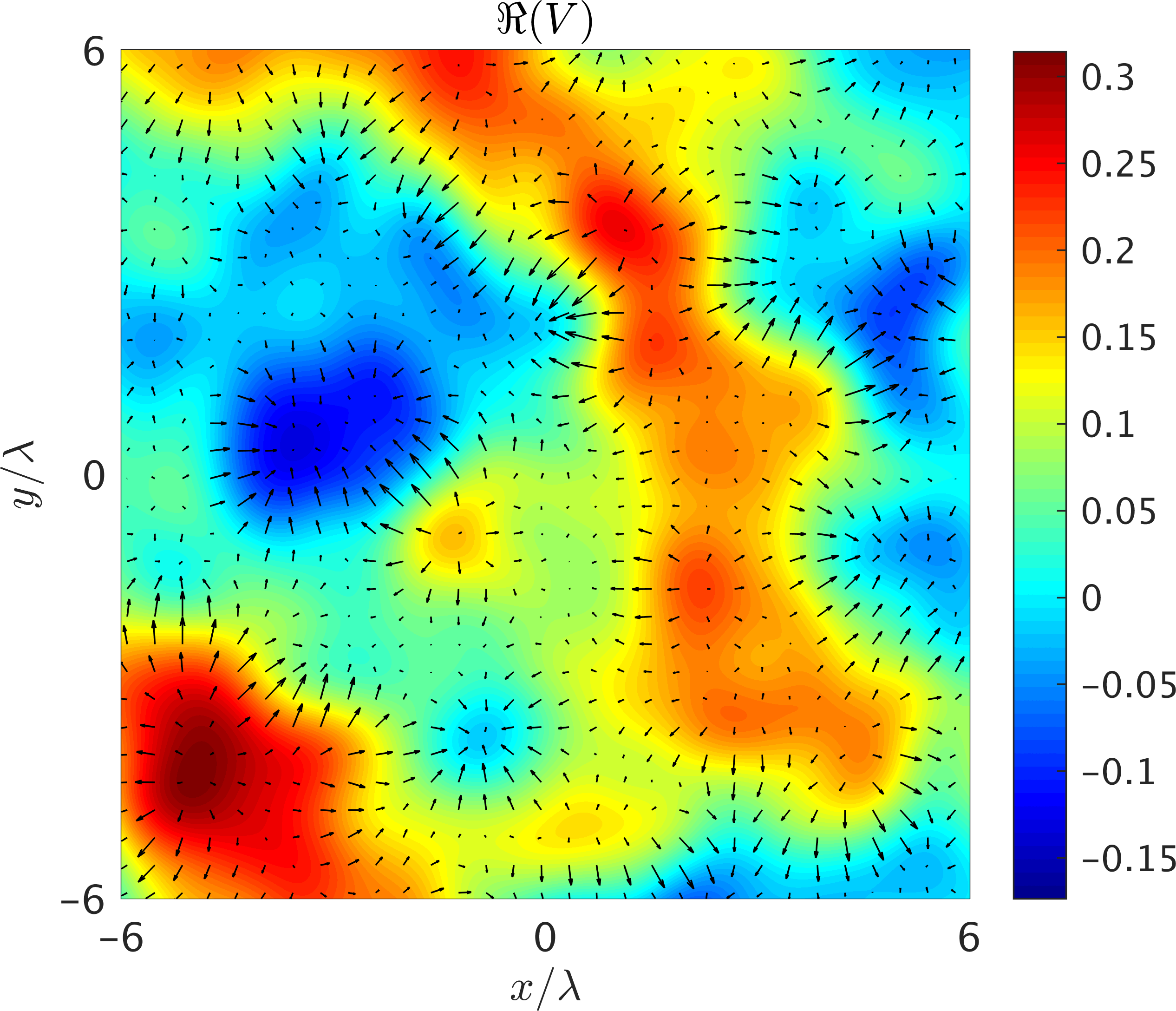}
 \includegraphics[width=.9\columnwidth]{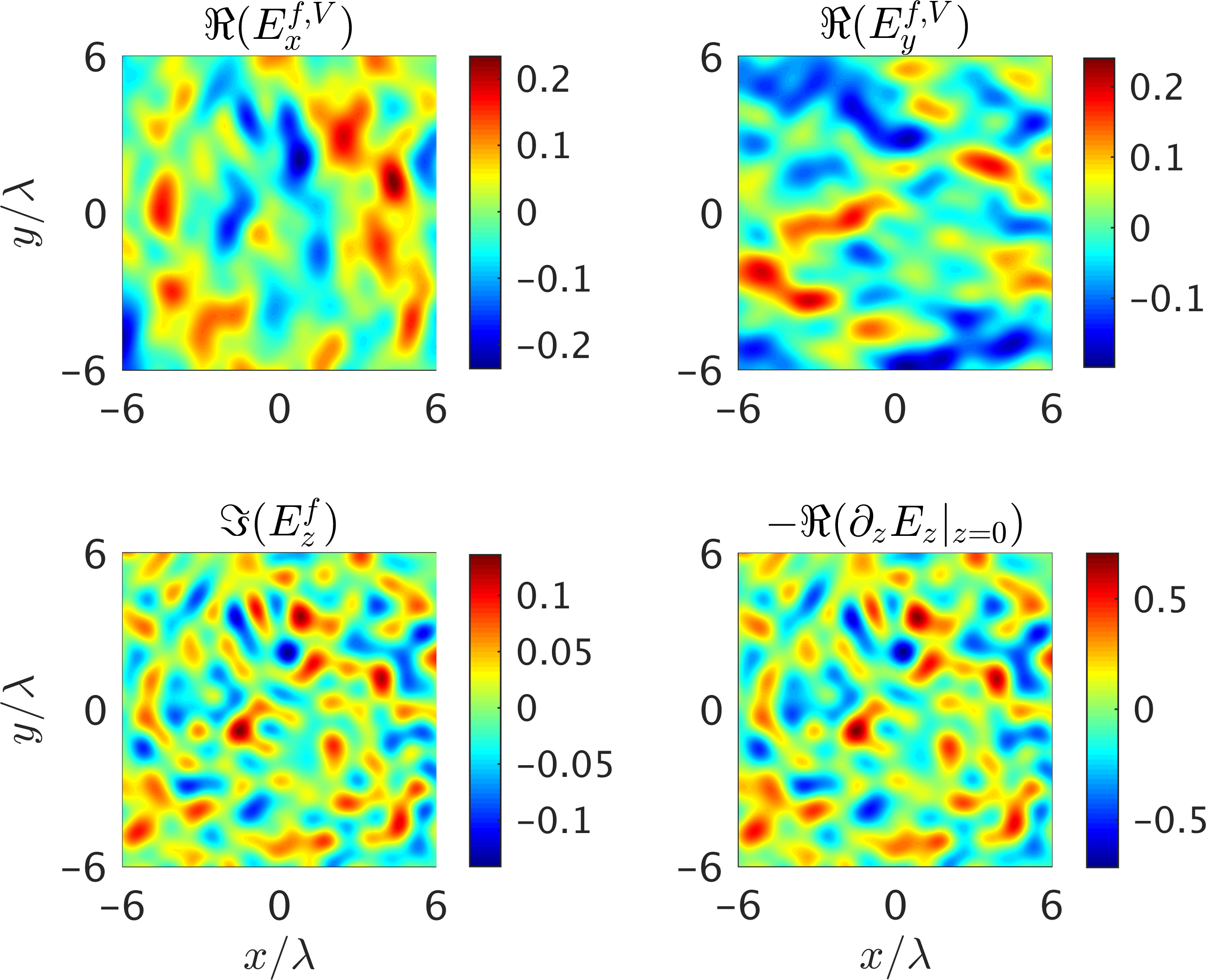}
 \caption{A random profile is prescribed in $\Im E_z^{\rm f}$ (see text for details). The real-valued "charge density" $-\partial_z E_z|_{z=0}$ looks very similar. Corresponding potential $V$ and transverse field components are shown as well.}
\label{fig:random}
\end{figure}

Let us prescribe a rather complicated longitudinal field component $E_z^{\rm f}$ --- a random beam, as shown in~\reffig{fig:random}. For this case, we simply consider a super-Gaussian beam profile in $E_z^{\rm f}$, $\exp(-r^{10}/(10\lambda)^{10})$, and multiply the latter with complex random numbers at each point of the numerical grid. The complex-valued random numbers are constructed as $f=\xi_1\exp(2\pi i\xi_2)$, with real-valued random numbers $\xi_{1,2}\in[0,1]$. The resulting beam profile is then multiplied in transverse Fourier space by both filter functions $H_{k_0}$ and $H_0$, as introduced above.  In order to facilitate comparison with previous figures, we choose to show $\Im E_z^{\rm f}$, such that all other derived quantities in~\reffig{fig:random} are real valued. Furthermore, we plot only the center part of the beam.

We now make several important remarks regarding~\reffig{fig:random}. First of all, the ``charge density'' $-\partial_z E_z|_{z=0}$, to once more take the electrostatic analogy, features a pattern very close to that prescribed in $\Im E_z^{\rm f}$. This means that even here the approximation~\refeq{eq:lowest_order} yields insight, and the longitudinal component $E_z^{\rm f}$ can be identified directly as a ``charge density.'' Moreover, ~\reffig{fig:random} shows very nicely how the landscape of the ``electrostatic potential'' $\Re V$ induced by $\Im E_z^{\rm f}$ ``shapes'' the transverse polarization components $\Re{\bf E}_{\perp}^{\rm f,V}$. In contrast to the previous examples, the peaks and troughs of $\Im E_z^{\rm f}$ cannot be directly related to the peaks and troughs of $\Re V$. This can be understood by thinking of $V$ as being a convolution between the Green's function of the two-dimensional Poisson equation (which is long-ranged) and the ''charge density": the convolution cannot resolve the delicate structure of the source term and simply smears it out. This also explains why the structures in the transverse field components are larger than those in $E_z^{\rm f}$.

\subsection{Linked trefoils in transverse and longitudinal components}\label{sec:knots}
We discussed the remaining degrees of freedom when fixing the longitudinal polarization component $E_z^{\rm f}$  in section \ref{sec:examples}.\ref{sec:vortex}, where they were exploited to simplify the transverse profiles. In principle, nothing prevents us from prescribing two field components, say $E_z^{\rm f}$ and $E_y^{\rm f}$, in the focal plane. This allows us to engineer both transverse and longitudinal components, and the potentials $V$ and $W$ determine the complete optical field via~\refeq{eq:full_solution}. 

Here, we will demonstrate the tremendous possibilities of creating customized vector beams by creating knotted vortex lines in the longitudinal and one transverse polarization component ($E_y$).  Apart from optics, knotted and linked vortex lines have also recently been studied in a range of different contexts, such as classical fluid dynamics~\cite{Irvine:nature:2013}, excitable media \cite{Maucher:PRL:2016, Maucher:PRE:2017,Maucher:arxiv:2018,Binysh:arxiv:2018}, nematic colloids \cite{Tkalec:science:2011,Martinez:NatMat:2014,Smalyukh:PNAS:2015} and superfluids such as Bose-Einstein condensates~\cite{Proment:PRE:2012, Irvine:NatPhys:2016}. In the present context, optical vortex knots~\cite{Dennis:Nature:2010,Maucher:PRL:2018,Sugic:arxiv:2018} can be used to imprint topological light structures onto the latter~\cite{Ruostekoski:PRA:2005,Maucher:NJP:2016}. 

\begin{figure}[ht]
 \includegraphics[width=.9\columnwidth]{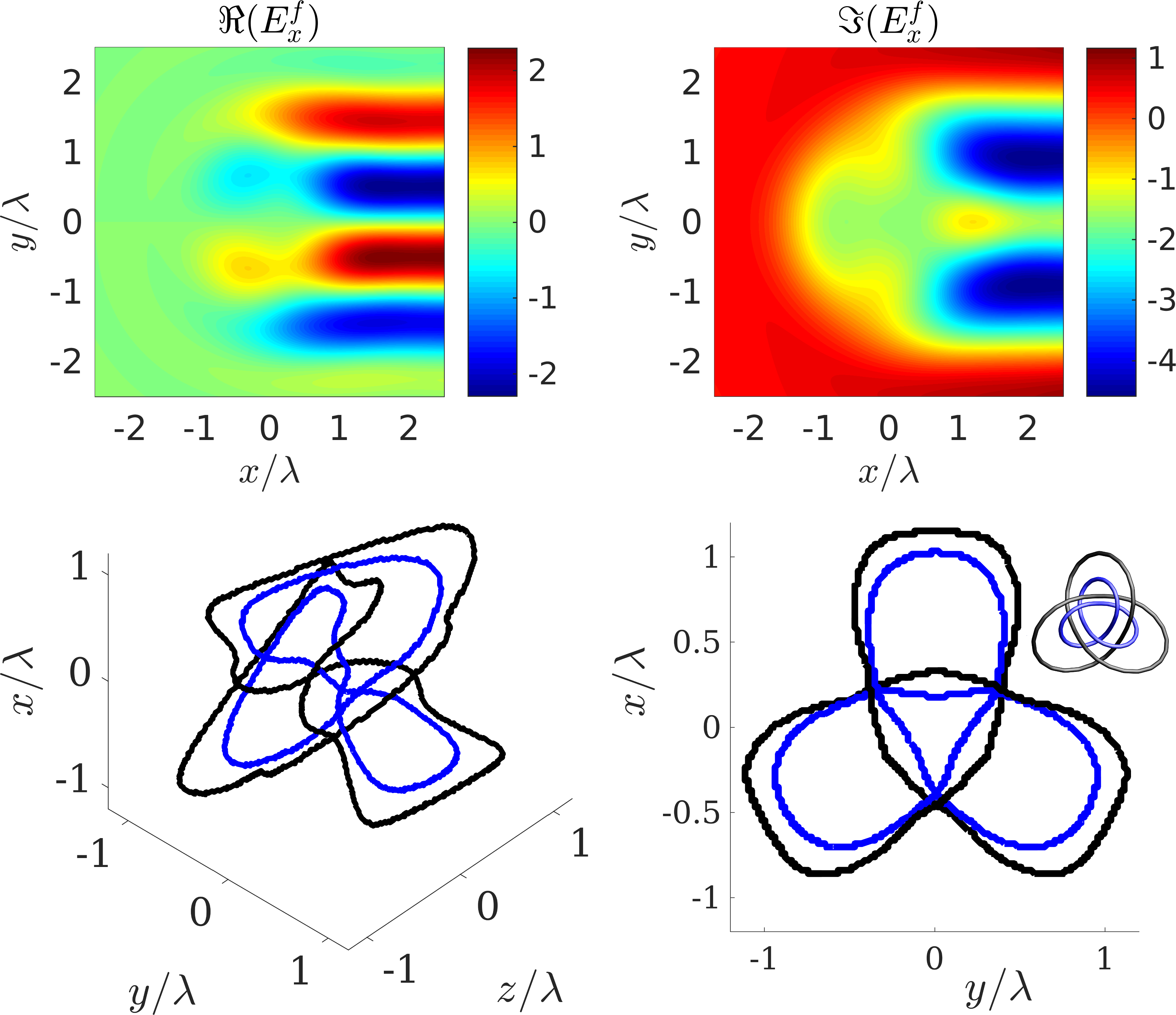}
 \caption{Two linked trefoils in different components of the electric field: The blue line represents the vortex line in $E_y$, the black line 
 the vortex line in $E_z$. The corresponding (semi-infinite) profiles in $E_x^{\rm f}$ are shown in the upper panels. 
 }
 \label{fig:double_trefoil}
\end{figure}

The topology of the two knots we envisage is sketched in~\reffig{fig:double_trefoil}: the vortex lines of the $E_y$ and $E_z$ components form two linked trefoil knots. To create such a vector beam, the respective components can be  prescribed in the focal plane as
\begin{equation}
\begin{split}
     E_y^{\rm f} &= 5  LG_{00}^{\sigma_y} - 7  LG_{01}^{\sigma_y} + 40  LG_{02}^{\sigma_y} - 18  LG_{03}^{\sigma_y} - 30  LG_{30}^{\sigma_y},\\
     E_z^{\rm f} &= 8  LG_{00}^{\sigma_z} - 18  LG_{01}^{\sigma_z} + 40  LG_{02}^{\sigma_z} - 18  LG_{03}^{\sigma_z} - 34  LG_{30}^{\sigma_z},
\end{split}
     \label{eq:trefoil_ez}
\end{equation}
with $\sigma_y=0.42\lambda$ and $\sigma_z=0.5\lambda$. The formulas of~\refeq{eq:trefoil_ez} were obtained according to~\cite{Dennis:Nature:2010,Maucher:NJP:2016}, with coefficients adapted to the non-paraxial situation~\cite{Maucher:PRL:2018}. We filter in transverse Fourier space with $H_{k_0}$ and $H_{0}$, where the latter is only necessary for $E_z^{\rm f}$.

Even though at first glance the two linked vortex knots in~\reffig{fig:double_trefoil} may seem somewhat contrived, they highlight that vortex lines in different polarization components can be chosen to be arbitrarily close without reconnecting. This is not possible for vortex lines in a single polarization component, which could be of importance when considering inscribing these vortex lines onto matter. Furthermore, such a topology demonstrates the tremendous possibilities of structured light. 

The transverse polarization component $E_x^{\rm f}$ is fully determined by~\refeq{eq:trefoil_ez}. Even though $E_z^{\rm f}$ and $E_y^{\rm f}$ have finite support, $E_x^{\rm f}$ is nonzero on a semi-infinite interval and thus impractical (see~\reffig{fig:double_trefoil}). Such semi-infinite field components occur when at least one of the other components integrated over the respective variable does not vanish: in our case, $\hat E_{y,z}^{\rm f}(k_x=0,k_y)\neq 0$. Nevertheless, this problem of impractical field components can be circumvented by simply attenuating the beam with, e.g., a sufficiently wide super-Gaussian profile, without affecting the propagation of the components of interest close to the optical axis. Additional satellite spots will appear far from the axis, but we have checked that for a super-Gaussian $\exp[-{\bf r}_{\perp}^{10}/(10\lambda)^{10}]$ those spots do not interfere with the linked vortex knots.

\section{Conclusions}

In this paper we have presented a systematic route to construct Maxwell-consistent vector beams. By means of a Helmholtz decomposition of the transverse polarization components in a plane transverse to the optical axis, we show that the full electromagnetic field can be generated by two scalar potentials $V$ and $W$. The potential $V$ produces ``curl-free'' (in the transverse plane) fields with a nonzero longitudinal component. The potential $W$ produces ``divergence-free'' (in the transverse plane) fields with zero longitudinal component. We suggest naming these \emph{generalized radial and azimuthal polarization states}, respectively. The decomposition of the optical field into generalized radial ($W=0$) and azimuthal ($V=0$) polarization states allows us to draw several analogies with other physical systems, i.e., electrostatics, magnetostatics, and fluid dynamics. By means of these analogies, it is possible develop an intuitive understanding of the interrelation between longitudinal and transverse polarization components, and the scalar potentials assume a ``physical meaning''. Finally, we presented several examples to illustrate the proposed decomposition and analogies. Besides rather simple configurations such as longitudinal vortices, we demonstrated arbitrary random beams as well as sophisticated topological light configurations. In all these examples, the above analogies were used to explain features in the respective polarization components, or even to conceive the beam configurations. 

We believe that our findings will broaden the range of accessible vector beams extensively and trigger further theoretical and experimental investigations involving structured light.

\section*{Funding Information}
This work is funded by the Leverhulme Trust Research Programme Grant RP2013-K-009, SPOCK: Scientific Properties Of Complex Knots.
S.S.\ acknowledges support by the Qatar National Research Fund through the National Priorities Research Program (Grant No.\ NPRP 8-246-1-060).

\bibliographystyle{unsrt}
\bibliography{bib}

\end{document}